\documentclass[acmsmall, natbib=true, review=false, anonymous=false, screen=true]{acmart}
\usepackage[]{graphicx}\usepackage[]{color}
\pdfoutput=1 
\makeatletter
\def\maxwidth{ %
  \ifdim\Gin@nat@width>\linewidth
    \linewidth
  \else
    \Gin@nat@width
  \fi
}
\makeatother

\definecolor{fgcolor}{rgb}{0.345, 0.345, 0.345}

\usepackage{framed}
\makeatletter
 {\par\unskip\endMakeFramed%
 \at@end@of@kframe}
\makeatother

\definecolor{shadecolor}{rgb}{.97, .97, .97}
\definecolor{messagecolor}{rgb}{0, 0, 0}
\definecolor{warningcolor}{rgb}{1, 0, 1}
\definecolor{errorcolor}{rgb}{1, 0, 0}
\newenvironment{knitrout}{}{} %

\usepackage{alltt}

\usepackage{acmcopyright}

\AtBeginDocument{%
  \providecommand\BibTeX{{%
    \normalfont B\kern-0.5em{\scshape i\kern-0.25em b}\kern-0.8em\TeX}}}

\setcopyright{rightsretained}
\acmJournal{PACMHCI}
\acmYear{2021} \acmVolume{5} \acmNumber{CSCW1} \acmArticle{56} \acmMonth{4} \acmPrice{}\acmDOI{10.1145/3449130}

\usepackage{subcaption}
\usepackage{tikz}
\usepackage{graphicx}
\usepackage{enumerate}
\usepackage{amsmath, amsthm} %

\usepackage[american]{babel}
\def\citepos#1{\citeauthor{#1}'s \cite{#1}}

\IfFileExists{upquote.sty}{\usepackage{upquote}}{}
\begin{document}

\newcommand{\mywidth}{.6\columnwidth}

\title[Effects of Algorithmic Flagging on Fairness]{Effects of Algorithmic Flagging on Fairness: Quasi-experimental Evidence from Wikipedia}

\author{Nathan TeBlunthuis}
\orcid{0000-0002-3333-5013}
\affiliation{%
  \institution{University of Washington}
\streetaddress{Box 353770}
\city{Seattle}
\state{Washington}
\country{USA}
\postcode{98195}
}
\email{nathante@uw.edu}
\authornote{Part of this author's contributions to this paper were made while he was affiliated with the Wikimedia Foundation.}

\author{Benjamin Mako Hill}
\orcid{0000-0001-8588-7429}
\affiliation{%
  \institution{University of Washington}
  \streetaddress{Box 353740}
  \city{Seattle}
  \state{Washington}
  \country{USA}
}
\email{makohill@uw.edu}

\author{Aaron Halfaker}
\orcid{0000-0001-8907-6367}
\affiliation{%
  \institution{Microsoft}
  \city{Redmond}
  \state{Washington}
  \country{USA}
}
\email{aaron.halfaker@gmail.com}
\authornote{The majority of this author's contributions to this paper was completed when he was affiliated with the Wikimedia Foundation.}

\authorsaddresses{Authors' adresses: Nathan TeBlunthuis, Department of Communication, University of Washington, Box 353740, Seattle, WA, 98195, USA; Benjamin Mako Hill, Department of Communication, University of Washington, Box 353740, Seattle, WA, 98195, USA; Aaron Halfaker, Microsoft, 1 Microsoft Way, Redmond, WA, 98052, USA}

\begin{abstract}
Online community moderators often rely on social signals such as whether or not a user has an account or a profile page as clues that users may cause problems. Reliance on these clues can lead to ``overprofiling'' bias when moderators focus on these signals but overlook the misbehavior of others. We propose that algorithmic flagging systems deployed to improve the efficiency of moderation work can also make moderation actions more fair to these users by reducing reliance on social signals and making norm violations by everyone else more visible. We analyze moderator behavior in Wikipedia as mediated by RCFilters, a system which displays social signals and algorithmic flags, and estimate the causal effect of being flagged on moderator actions.  We show that algorithmically flagged edits are reverted more often, especially those by established editors with positive social signals, and that flagging decreases the likelihood that moderation actions will be undone. Our results suggest that algorithmic flagging systems can lead to increased fairness in some contexts but that the relationship is complex and contingent.
\end{abstract}



\maketitle

\section{Introduction}

Online community moderators are responsible for reviewing the torrents of user-generated content for spam, vandalism, attacks, and other violations of community norms and rules.  In many large online communities, a small number of moderators---often volunteers---will be responsible for reviewing thousands or millions of actions and taking steps to stop and mitigate problematic behaviors \cite{gillespie_custodians_2018}. To help focus their attention within this deluge, moderators typically rely on social signals \cite{donath_social_2014} that indicate that a user's contributions are made in good faith and of high quality \citep{kraut_building_2012}. Common signals include visible reputation scores, user profiles, experience, and registration status \cite{broughton_wikipedia_2008, kraut_building_2012}. 
For example, since new users are often more likely to engage in bad behaviors, moderators might scrutinize contributions from newcomers more closely \citep{kraut_building_2012, potthast_automatic_2008}.
However, directing limited moderation attention based on social signals can introduce unfairness through ``overprofiling'' that occurs when moderators focus their attention on users with signals associated with bad behaviors while ignoring others engaged in similar or worse behaviors \cite{de_laat_profiling_2016}. 
For this reason, and because relying on social signals can still place enormous demands on limited moderator resources, online communities are increasingly adopting algorithmic flagging systems to direct moderators toward problematic actions \cite{chandrasekharan_crossmod:_2019, halfaker_ores:_2020}.

Although the consequences are very different, these systems share salient commonalities with algorithmic flagging systems used in employment, college admissions, and criminal justice. All of these systems use predictions of whether an outcome will occur to flag certain individuals as more or less likely sources of problems. All leave final decisions to a human judge.
The use of these systems when people's lives are at stake has rightfully attracted criticism based of how algorithms engage in misrepresentation and discrimination \cite{campolo_ai_2017, oneil_weapons_2018,barocas_fairness_2019}. 
On the other hand, advocates of algorithmic prediction in criminal justice argue that algorithms---even those that are measurably biased in their predictions---might still be less discriminatory than decisions made by biased human judges alone \cite{kleinberg_human_2018, stevenson_assessing_2017}.

Can algorithmic flagging systems in online community moderation similarly reduce reliance on social signals and lead to more fair outcomes? We aim to answer this question through a field evaluation of an algorithmic flagging system called RCFilters, which was deployed on 23 different Wikipedia language editions from January 2019 to March 2020.  RCFilters flags contributions identified by the Objective Revision Evaluation Service (ORES) machine learning system as likely to be damaging \citep{halfaker_ores:_2020}. These flags are shown along with existing social signals of quality. We take advantage of a set of arbitrary thresholds built into RCFilters to conduct a quasi-experimental analysis that estimates the causal effect of algorithmic flagging on moderation decisions and that seeks to measure whether algorithmic flags lead to better or worse outcomes for users who are likely to be overscrutinized \textit{ex ante}.
Our results suggest that algorithmic flagging can lead to more fair outcomes but that this effect may depend on the specifics of the social signals in question.

Our paper makes several contributions.
First, our work answers calls to analyze the impacts of algorithms \textit{in situ} \cite{selbst_fairness_2019, stevenson_assessing_2017, zhu_value-sensitive_2018} by offering an empirical evaluation of an algorithmic flagging system in an important social computing context. 
Second, our analysis contributes to an ongoing debate over when and how algorithms might lead to more or less fair outcomes for individuals subject to profiling by human decision makers.
Third, our work offers a methodological contribution by presenting a novel quasi-experimental approach that can act as a template for future non-interventional studies of causal effects of algorithmic decision support systems.
Finally, our work contributes to social computing system design by suggesting improvements to algorithmic flagging and filtering systems.

\section{Background}

\subsection{Moderation in Online Communities} 
Contemporary online communities are flooded with harassment, spam, misinformation, disinformation, and hate. Users of social media systems frequently and flagrantly violate community and platform rules, various laws, and norms of decency and decorum. Even users acting in good faith can do damage by taking conversations off-topic, undermining the stated purpose of communities, and lowering the quality of discourse or the knowledge goods being produced. Protecting online communities from unwanted activity are content moderators---many of them volunteers---that \citet{gillespie_custodians_2018} has described as ``custodians of the Internet.''
Moderation work typically involves three tasks: namely, reviewing content or activity, mitigating damage caused by a problematic behavior, and sanctioning users in different ways \citep{gillespie_custodians_2018, seering_moderator_2019, jiang_moderation_2019, kiene_technological_2019}.

\citet{grimmelmann_virtues_2015} defined moderation as ``governance mechanisms that structure participation in a community to facilitate cooperation and prevent abuse.'' %
Discussions of content moderation often focus on individuals occupying formal roles as moderators with special rights and responsibilities. For example, several of the moderators in \citepos{gillespie_custodians_2018} account are professional moderators working for major platforms such as Facebook and Twitter. Several moderators, and nearly all of them on platforms such as Reddit and Discord \citep{matias_going_2016, jiang_moderation_2019, kiene_technological_2019}, work as volunteers but occupy similar positions of formal authority and responsibility.
That said, the work of moderation is also distributed across regular community members \cite{lampe_slashdot_2004, kiene_surviving_2016}. In Wikipedia, for example, the bulk of moderation activity as defined by \citeauthor{grimmelmann_virtues_2015} occurs as normal users review, vet and undo the work of others to mitigate damage and sanction users they believe have behaved badly \citep{piskorski_testing_2017}.

\subsubsection{Sanctions}

Sanctioning involves enforcing norms in ways that attempt to discourage future misbehavior. It is a core part of moderation work because it encourages compliance with norms by communicating that rules will be enforced \cite{jhaver_did_2019, srinivasan_content_2019}. Although it also serves to mitigate damage, removing content is a common form of sanctioning because it communicates that an action was inappropriate \citep{piskorski_testing_2017}. \citet{halfaker_dont_2011} showed that removing content is an effective sanction and results in higher quality contributions by the reverted contributor in Wikipedia. Similarly, \citet{srinivasan_content_2019} found that people whose comments were removed from Reddit were less likely to violate norms in the future.

Although the goal of most sanctioning is to steer participants toward productive behaviors, the effect is often to deter participation. This can be particularly problematic with well-meaning newcomers who often violate norms because they have not yet learned the ropes \cite{adler_content-driven_2007, halfaker_dont_2011, halfaker_rise_2013}.
Sanctioned newcomers are less likely to continue participating, especially in the absence of clear explanations from moderators \cite{jhaver_did_2019, kraut_regulating_2012, potthast_automatic_2008, halfaker_rise_2013, teblunthuis_revisiting_2018}.
On Wikipedia and similar communities, high rates of sanctioning can help explain declines in participation and may be an obstacle to building a community that includes diverse participants \cite{halfaker_rise_2013, teblunthuis_revisiting_2018, lam_wp:clubhouse?:_2011}.

\subsubsection{Meta-norms}

No moderation system is perfect. Moderators inevitably make mistakes and apply sanctions in ways that are arbitrary and unfair. This is particularly challenging to avoid in distributed moderation models used on sites such as Slashdot or Wikipedia where moderation is conducted  by large and diverse groups of untrained and loosely coordinated users.
Sanctions can be particularly demotivating to newcomers who feel that sanctions are unfair and incorrect \citep{srinivasan_content_2019, jhaver_did_2019, gillespie_custodians_2018}.
Consequently, steps that make moderation more fair might decrease the negative effects of sanctions on community growth.

One way to improve fairness in moderation is through governance structures that enforce accountability \citep{frey_this_2019}.
Toward this end, Slashdot famously created tools for ``meta-moderation'' that allowed all users to evaluate the decisions of moderators \cite{lampe_slashdot_2004}. Users whose moderation decisions were controversial or at odds with the opinions of other Slashdot members would not be given moderation privileges again.
Although formal systems for meta-moderation remain rare, behaviors that take action against controversial sanctions are common and serve a similar social function \citep{crawford_what_2016}.    
``Meta-norms,'' which prescribe when and how one should issue sanctions against violations of first-order norms \cite{horne_enforcement_2001} are particularly relevant.  \citet{reagle_be_2010} documented the formalization of meta-norms on Wikipedia and \citet{piskorski_testing_2017} showed how Wikipedia users maintain meta-norms by undoing sanctions in ways that effectively sanction the originally sanctioning user.

\subsubsection{Flagging and Algorithmic Triage}

Moderators in large online communities can face incredible challenges in scaling their work to handle an enormous mass of content and user activity  \citep{gillespie_custodians_2018, kiene_technological_2019, seering_moderator_2019, seering_shaping_2017}. In interviews conducted by \citet{kiene_technological_2019}, small teams of volunteers tasked with maintaining order in large communities described their work as akin to ``running a small city.'' 
Some platforms deal with scale by employing more paid moderators. However, the work involved can be exploitative, challenging, traumatizing, and expensive \cite{roberts_commercial_2016}. Volunteer moderator teams frequently find it difficult to identify, train, and integrate new members as they grow \citep{kiene_surviving_2016}. On average, teams become less likely to add new members as their communities grow \citep{shaw_laboratories_2014}. 

For these reasons and others, it is often impossible for communities to scale moderation resources such that human moderators can review all activity.
As a result, many moderation systems implement flagging so that a wider group of users can report content for review by moderators \cite{grimmelmann_virtues_2015}.
If users reliably flag problematic behaviors, flagging can mitigate issues of scale because moderators focus their attention on behavior that is flagged. Obviously, flagging is far from a perfect solution. 
From the perspective of a flagged user, flagging can seem arbitrary and opaque \cite{crawford_what_2016}.  
From a moderator perspective, flagging is flawed because disgruntled users can coordinate to overwhelm moderators and target opposing viewpoints \cite{crawford_what_2016}. 
Finally, given that traditional flagging systems continue to rely on volunteer labor, they often fail to fully address issues of scale, leaving many bad actions unflagged, unreviewed, and unsanctioned. 

To address this final limitation, communities have turned to algorithmic flagging systems that use computer programs to automatically mark content for review by human moderators \citep{kiene_technological_2019,kiene_who_2020,seering_shaping_2017}. Although some of these systems rely on keywords, regular expressions, or heuristics, more advanced and flexible versions of these systems use predictions from machine learning models. These systems are seen as promising answers to the problem of moderation at scale because they can be easily be used to review an enormous volume of behaviors, they may be less vulnerable to strategic flagging, and they may be more reliable than human reviewers.

Algorithmic flagging systems can be thought of as human-in-the-loop versions of similar computational systems that engage in fullly automated moderation. For example, numerous digital platforms utilize the PhotoDNA system to automatically identify and remove child pornography \cite{gillespie_custodians_2018}. Similarly, Wikipedia's ClueBot NG uses a machine learning predictor to automatically remove vandalism \cite{geiger_when_2013}. Although they play a critical role in reducing moderation workloads, fully automated systems are uncertain enough in most of their assessments that they are typically only considered useful in defending against the most clear-cut examples of misbehavior \cite{gillespie_custodians_2018}.

Some machine learning systems that are designed to classify bad behavior are used as a form of algorithmic triage. While the most egregious examples of bad behavior are dealt with by automatic systems, other possible norm violations are flagged for review by human moderators.
For example, Reddit allows moderators to define a system of rules based on regular expressions to automatically flag content for further review \cite{jhaver_human-machine_2019}. Algorithmic flagging systems based on machine learning occupy the vanguard of online activity regulation and numerous examples have been described in recent scholarship. 
\citet{chandrasekharan_crossmod:_2019} described a system for Reddit communities to share information and collaborate on automatic flagging that accounts for differences between rules of different communities.
\citet{wulczyn_ex_2017} presented a system for classifying harassing behavior on Wikipedia. Finally, \citet{halfaker_ores:_2020} developed the ORES system to predict the quality of contributions and content on Wikipedia.

\subsection{Will Algorithmic Flagging Decrease Discrimination Of Overprofiled Users?}

One of the most important debates in contemporary technology policy is the degree to which the introduction of algorithms into socially consequential decision making leads to more or less fair outcomes \citep{chouldechova_fair_2017, kleinberg_human_2018, oneil_weapons_2018, selbst_fairness_2019}. Much of this debate focuses on arguments about whether algorithms will amplify or entrench discrimination and on biases introduced by training data \cite{barocas_fairness_2019, campolo_ai_2017, sap_risk_2019}.
Discrimination is the deferential treatment of individuals based on membership in a group. Economists of discrimination distinguish between taste-based and statistical discrimination \citep{becker_economics_1957, bertrand_field_2016, phelps_statistical_1972}.  Taste-based discrimination is driven by preferences for members of one group and includes both ideologically-driven racism and implicit bias.  Statistical discrimination occurs when social signals---visible and socially salient characteristics, such as group memberships---are instrumental in driving decisions. Statistical discrimination can also lead to unequal outcomes for certain groups.

\subsubsection{Social Signals}
\label{sec:social.signals}

Although most discussions of discrimination focus on high-stakes contexts such as banking, labor markets, and criminal justice, moderation in online communities is also ripe for statistical discrimination based on visible social signals.
When interacting in face-to-face groups, people can observe---and discriminate on the basis of---visible signals of status, group membership, psychological states, or cultural identity \cite{donath_social_2014, pentland_honest_2008, ridgeway_status:_2019}. Because the invisibility of these signals in online communities creates a barrier to regulation, sociability, and cooperation, 
communities use devices such as profile images and biographies, avatars, or visualizations of activity as tools for self-presentation and signals of membership \cite{lampe_familiar_2007, donath_social_2014}.  
Disclosing information on profiles can provide signals helpful for people using prototypes \cite{grabner-krauter_trust_2015}, building social capital \cite{ellison_connection_2011}, and developing trust \cite{ma_self-disclosure_2017}. Formal reputation systems such as karma on Reddit and Slashdot or badges on StackExchange can be important signals of commitment, quality, and trustworthiness \cite{grimmelmann_virtues_2015, lampe_role_2012, merchant_signals_2019}

Even without user profiles or formal reputation systems, participants in online communities use subtle signals to draw conclusions about each other \cite{donath_signals_2007, ellison_managing_2006, jacobson_impression_1999}. Sparse cues such as usernames or communication styles can be signals of personality, gender, and identity \cite{donath_social_2014, hancock_impression_2001, herring_gender_2000}. Tests of community-specific technical or cultural knowledge can identify newcomers and, similar to formal reputation systems, they may be more challenging to fake than biographical information \cite{bernstein_4chan_2011, donath_social_2014, grimmelmann_virtues_2015}.
In peer production projects, prior contributions can be inspected for information about expertise, work styles, and the future value of a newcomer \cite{marlow_impression_2013}. 

In several online communities such as Wikipedia, users can elect to participate anonymously, under more-or-less stable pseudonyms, or using their real names. Masking signals of gender, race, age, (dis)ability, or status can appear to equalize and free individuals from oppressive prejudices and stereotypes \cite{dubrovsky_equalization_1991, friedman_social_2001}.
On the other hand, the presence or absence of a stable user identity is itself an essential signal because persistent identities make it possible to build up reputation, social capital, and trust and the inability to do so is associated with misbehavior \cite{grabner-krauter_trust_2015, hill_hidden_2020}.

\subsubsection{Will algorithmic flagging reduce overprofiling?}

Online community moderators can use social signals to discover and respond to misbehavior, but this can lead to statistical discrimination.
Wikipedia's \textit{Missing Manual} advises would-be vandal fighters on Wikipedia to ``consider the source'' when ``estimating the likelihood that an edit is vandalism'' \cite{broughton_wikipedia_2008}.
Because newcomers and anonymous users are more likely to violate rules, moderators may rely on social signals of newness to find bad behaviors or to decide if an ambiguous contribution was made in bad faith.
Increased scrutiny and skepticism can translate into an increased likelihood of sanction, simply for being new or anonymous.  
Statistical discrimination emerges because moderators are more likely to scrutinize and sanction new or anonymous contributors who have legitimate reasons for contributing. 

Ethical philosophers have objected to the way social signals are used in online moderation activity. Dutch philosopher Paul de Laat adopted the concept of ``profiling'' from legal scholar Frederick Schauer to argue against the use---and even the public display of---social signals such as registration status and experience levels in the user interfaces used for moderation because they are prone to ``overuse'' \cite{de_laat_use_2015, de_laat_profiling_2016}. It should be noted that discriminating by attributes such as newness does not raise the same legal or constitutional concerns as discrimination against protected classes such as race or religion.  Online communities establish their own norms and may choose to protect or target certain attributes on the basis of a specific community's values. 
For example, while discussing Wikipedia, de Laat argues that ``overuse'' is unethical, immoral, and inconsistent with the community's founding principles of transparency and equality. Drawing on de Laat, we refer to individuals with social signals that elicit undue scrutiny as ``overprofiled.''

Although an important debate continues over the use of algorithmic predictions in domains like criminal sentencing, proponents of algorithms argue that they could reduce discrimination and inequality \cite{kleinberg_human_2018, stevenson_assessing_2017}. Algorithms can reproduce statistical discrimination, but they might be less biased than the alternative: human decisions that would presumably rely heavily, if perhaps subconsciously, on salient social signals such as race. Critics suggest that algorithms simply obscure this discrimination behind complex mathematical models that are difficult to understand, interrogate, or challenge.

Although this debate is challenging to resolve in the case of criminal justice, algorithmic flagging in online community moderation provides a setting with lower stakes and more detailed data. 
If we apply arguments proposing that algorithms can reduce discrimination to community moderation, we would conclude that algorithmic triage systems would reduce the impact of discrimination among overprofiled individuals by making misbehavior by all kinds of users visible to moderators. If algorithmic flagging reduces overprofiling bias, then it will have a smaller effect on overprofiled users than on others. If algorithms simply reproduce discrimination, we would find no such difference. 
This leads us to our first research question: 
\textit{\textbf{[RQ1]}  How will flagging an action change the likelihood an action is sanctioned for overprofiled editors compared with others?}

Algorithmic fairness researchers use specific criteria to quantify biases encoded in algorithmic predictors and the fairness of resulting decisions \cite{chouldechova_fair_2017, barocas_fairness_2019, mitchell_prediction-based_2020}. 
These criteria are often developed for settings where model predictions are equivalent to decisions. For example, \citet{kusner_counterfactual_2017} define demographic parity in terms of model predictions, whereas \citet{mitchell_prediction-based_2020} define it in terms of human decisions. In algorithmic flagging, decisions are informed by algorithms but left to humans.  
Therefore, we distinguish between the fairness of predictions and the fairness of decisions and refer to our criteria as ``decision system fairness metrics'' following \citepos{mitchell_prediction-based_2020} use of the term ``decision system.''

We first consider demographic parity, as shown in Equation \ref{eq:demographic.parity}, which means that the probability of a decision ($D$) is statistically independent of a protected attribute ($A$) \cite{kusner_counterfactual_2017,  barocas_fairness_2019}:

\begin{equation}
    P(\widehat{D} \vert A = 0) = P(\widehat{D} \vert A = 1)
\label{eq:demographic.parity}
\end{equation}
\noindent An algorithmic flagging system will have demographic parity concerning registration status if the probability that an action is flagged is the same for actions by overprofiled and underprofiled editors.
Our analysis of RQ1 thus evaluates how flagging shapes demographic parity for sanctioning decisions.

\subsection{Will Algorithmic Flagging Increase Fairness?}

A system might lack demographic parity by sanctioning one group more than others but still be justifiable if all sanctions are fair.
What does it mean for a sanction to be fair? The subject of fairness in algorithmic systems is a major subject of debate in computing and AI. There are several different approaches to conceptualizing fairness, and no algorithmic predictor can satisfy them all \citep{barocas_fairness_2019, caraban_23_2019, kleinberg_inherent_2016,  mitchell_prediction-based_2020, yin_understanding_2019,wallach_big_2019}. 
While such approaches focus on discrimination built into machine learning programs, we seek a concept of fairness that reflects the standards of relevant communities of practice.  We find one in the concept of ``meta-norms'' from social psychology and James Coleman's sociological conception of norm maintenance. Drawing from these sources, we define unfair sanctions as those that a community is unwilling to let stand---i.e., sanctions that are themselves the subject of sanction \citep{coleman_social_1988, horne_enforcement_2001, piskorski_testing_2017}. 
For example, norms in Wikipedia govern right and wrong ways of editing wiki pages. Sanctions of first-order norm violations are governed by meta-norms about what sorts of contributions merit sanction. Following \citet{piskorski_testing_2017}, we describe a sanction as \emph{controversial}---i.e., in likely violation of a meta-norm---if it, in turn, is sanctioned by a third community member.

A controversial sanction suggests that the initial edit was not truly damaging (i.e., $D=1$ but $Y=0$ where $Y=1$ means an edit was truly damaging). Thus, a controversial sanction is analogous to false positive classification by a machine predictor ($ \widehat{Y}=1$ but $Y=0$, where $\widehat{Y}=1$ means the machine predicts that an edit is damaging). The false positive rate quantifies the amount of unfair treatment a group experiences,  but it does not compare unfair treatment between groups. Therefore, is not strictly speaking an algorithmic fairness criterion. However, changes in the false positive rate of the decision system (shown in Equation \ref{eq:change.sanction.fpr}) quantify how flagging is increasing or decreasing the rate of unfair sanctions.

\begin{equation}
    P(D=1 \vert Y=0, \widehat{Y}=1) - P(D=1 \vert Y=0, \widehat{Y}=0)
\label{eq:change.sanction.fpr}
\end{equation}

\noindent Relying on this definition of fairness, our second research question asks how algorithmic flagging shapes the fairness of sanctioning in terms of the rate of sanctions for meta-norm violations: \textit{\textbf{[RQ2]} How will flagging an action change the chances it receives a controversial sanction?}

Influential theoretical frameworks in social computing seem to predict competing answers to this second question. 
First, dual-process models of behavioral economics suggest that people will tend to rely on ``salient signals'' for rapid decision making in conditions of uncertainty and imperfect information \citep{bordalo_salience_2012, kleinberg_human_2018, tversky_judgment_1974}.  When human moderators use social signals to choose behavior to review or sanction, these attributes serve as salient signals but remain far from perfect signals of quality.
Algorithmic flags provide an additional salient signal but are also far from perfect \cite{halfaker_ores:_2020}.  Indeed, algorithmic flagging systems are typically designed to minimize the risk of missing bad behaviors by surfacing large numbers of false positives (i.e., non-problematic behaviors) and relying on human moderators to make final decisions.
Of course, if human moderators use algorithmic flags as salient signals, they may reproduce algorithms' false predictions. In this case, controversial sanctions will increase.

A second perspective suggests that algorithmic flags can increase fairness.
Several online communities have institutionalized rules, norms and meta-norms and act as highly bureaucratic organizations \cite{butler_dont_2008, piskorski_testing_2017}. 
Max Weber described how bureaucratic organizations construct and use two concepts of what he called ``rationality:'' substantive rationality and formal rationality  \cite{weber_economy_1978}.  
Substantive rationality refers to how bureaucratic organizations use policies, routines and hierarchy to define their collective values and goals. 
Formal rationality refers to the use of calculated decision making, such as that involving productivity or financial metrics, in the pursuit of goals \cite{lindebaum_insights_2019}.  
Following Weber, \citet{kreiss_limits_2011} argued that increasing substantive rationality through bureaucratic policies in online communities can lead to more fair outcomes.  

Although less explored by scholars of online communities, there are also reasons to believe that increasing formal rationality in moderation decisions might also enhance fairness, at least in online communities with mature normative systems. In such contexts, algorithmic flagging systems can enact formal rationality by estimating the probability and displaying an authoritative signal that an action runs afoul of shared behavioral standards. Adopting algorithmic flagging can thus mark a shift away from idiosyncratic individual decision-making and toward increasing the use of formalized rationality. Through this lens, an algorithmic flagging system---even one that encodes biases---can be a ``carrier of formal rationality'' \cite{lindebaum_insights_2019}, leading to governance that is more in line with community meta-norms and to a decrease in controversial sanctions. 

Next, we consider how changes in the false positive rate of the decision system depends on overprofiling. This corresponds to evaluating decision system fairness in terms of equality of opportunity (shown in Equation \ref{eq:balance}) \cite{hardt_equality_2016, mitchell_prediction-based_2020}:
\begin{equation}
    P(D = 1 \vert Y = 0, A=0) = P(D = 1 \vert Y=0, A=1)
\label{eq:balance}
\end{equation}

\noindent Equality of opportunity is satisfied when the false positive rate of a decision system does not depend on the protected attribute.  
Equality of opportunity for registration status would mean that registered and unregistered editors that make good edits have equal chances of having their contributions accepted.

Our third research question asks whether algorithmic flagging systems will increase or decrease equality of opportunity: \textit{\textbf{[RQ3]} Within the set of sanctioned actions, how will the effect of flagging an action on controversial sanctions depend on whether contributors are overprofiled?}
                  
Once again, influential theoretical frameworks in social computing research seem to point in opposite directions.  Under dual-process psychological models, both social signals and algorithmic flags might cue moderators to issue sanctions and might substitute for one another. In this case, we would hypothesize that flagging would have a more positive effect on controversial sanctions among underprofiled contributors, who had previously been relatively ignored, than it does among the overprofiled individuals, who were always scrutinized.
Conversely, if the larger effect of algorithmic flagging is helping moderators comply with meta-norms, it simply will not matter whether contributors are overprofiled.

\section{Empirical Setting}
\label{sec:empirical}

\begin{figure}[t]
  \centering
\begin{tikzpicture}

  \node[anchor=west](flags) at (-7,2.7) {ORES Flags};
  \node[anchor=west](userpagelink) at (-1.7,2.7) {User profile link};
  \node[anchor=west](unregistered) at (1.4,2.7) {Unregistered editor};

  \begin{scope}
  \node[anchor=south, inner sep=0] (image) at (0,0) {\includegraphics[width=\textwidth]{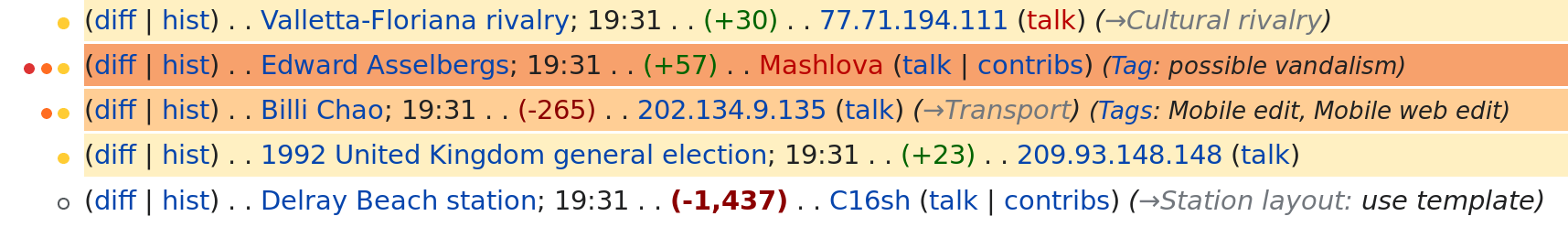}};
  \draw [-stealth,ultra thick] (flags.220) -- ++(0,-0.4);
  \draw [-stealth,ultra thick] (unregistered.200) -- (1.5,2);
  \draw [-stealth,ultra thick] (userpagelink.280) -- (0.2,1.6);
\end{scope}
\end{tikzpicture}

  \caption[Screenshot of edit metadata shown in RCFilters.]{Screenshot of Wikipedia edit metadata on Special:RecentChanges with RCFilters enabled.  Highlighted edits with a colored circle to the left side of other metadata are flagged by ORES.  Different circles and highlight colors (white, yellow, orange and red in the figure) correspond to different levels of confidence that the edit is damaging. Users can configure which colors are shown.  Visible social signals include registration status (i.e., whether a username or an IP address is shown) and whether an editor's user page and user talk page exist.  RCFilters does not specifically flag edits by new accounts, but does support filtering changes by newcomers.}
  \label{fig:rcfilters}
\end{figure}
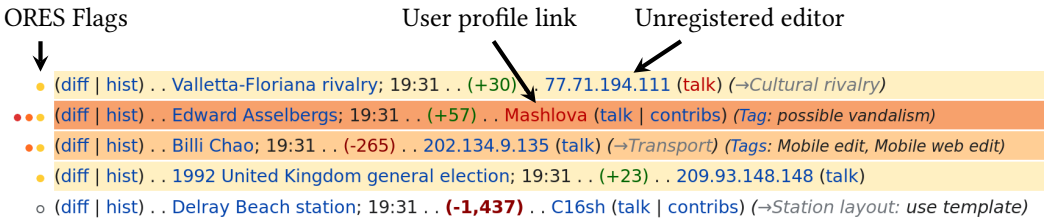

We aim to answer our three research questions through a field evaluation of an algorithmic flagging system called RCFilters, which was deployed on  23 different Wikipedia language editions between January 2019 and March 2020. RCFilters stands for ``Recent Changes filters.'' The term ``Recent Changes'' refers to a page on Wikipedia that allows viewers to see the most recent changes made to the site.\footnote{For example, the Recent Changes page for English Wikipedia is available here: \url{https://en.wikipedia.org/wiki/Special:RecentChanges} (Archived: \url{https://perma.cc/BNZ3-E9D5})} As Figure \ref{fig:rcfilters} shows, RCFilters adds a set of flags represented as colored dots on the left side of the list of recent contributions. Social signals are also visible, including registration status and whether a user has created a profile page.  Although dense with information regarding recent edits and hyperlinks, the page is immediately understandable to Wikipedia moderators. When deployed, the RCFilters interface appears both on ``Recent Changes'' as well as on  ``watchlists''---a special version of ``Recent Changes'' that shows only edits to the subset of pages that a user has elected to follow. RCFilters must be enabled by each user on their Wikipedia user preferences page.

Algorithmic flagging in the RCFilters system is powered by the ORES edit quality models trained to predict whether edits are labeled ``damaging'' or ``not damaging.'' The models are gradient boosted decision trees trained on a mixture of human-labeled Wikipedia edits and edits made by established editors that are assumed to be ``not damaging.''  

It should be noted that ORES models do not merely reproduce profiling patterns typical of moderation on Wikipedia.  
The interface for labeling training data obscures social signals from the volunteer Wikipedians doing labeling work and its models are predictive of damage from users that are not anonymous or newcomers. 
Nevertheless, as discussed in §\ref{sec:threats}, ORES encodes biases against unregistered editors and---to a lesser extent---against editors without user pages.
ORES was designed neither to merely support quality control in Wikipedia, nor to optimize precision, recall, or fairness but to enact Wikipedian principles of openness, transparency, and community accountability---to ``deploy efficient machine learning at scale for content moderation \ldots\ in ways that enable volunteers to develop and deploy advanced technologies on their own terms'' \citep{halfaker_ores:_2020}. More information on the philosophy, design and implementation of ORES can be found in \citeauthor{halfaker_ores:_2020} \cite{halfaker_ores:_2020}.

\section{Methods}

Our analysis is based on a regression discontinuity design (RDD) that aims to estimate causal the effects of flagging by RCFilters on moderator behavior in Wikipedia \cite{imbens_regression_2008, jacob_practical_2012, lee_regression_2010}. Common in empirical economics, RDDs are quasi-experimental in that they resemble a randomized control trial for data points in the neighborhood of an arbitrary cutoff \cite{jacob_practical_2012, lee_regression_2010}. 
RDDs model how an outcome depends on this cutoff and a continuous ``forcing variable.'' The idea behind an RDD is that observations immediately below and above the cutoff will be equal in expectation after adjusting for any underlying (i.e., ``secular'') trend. For example, RDDs used in econometrics might estimate the effect of passing a test by comparing the outcomes of people who barely passed and failed. 
One benefit of an RDD over a field experiment based on A/B tests is that it can provide ecological validity and support causal claims without subjecting users to intervention without consent \citep{lane_big_2015, jouhki_facebooks_2016}. 
Although they remain rare in computing, RDDs have been used in recent publications in social computing \citep{narayan_all_2019, hill_hidden_2020}.

Our forcing variable is the score from the ORES machine learning system. Our cut-off variables are a set of arbitrarily chosen operating points used by RCFilters. Our outcomes are constructed by creating two variables that indicate whether a revision's author is overprofiled as well as variables that indicate whether each revision was reverted or subject to a controversial revert. We discuss each in turn before introducing our analytic approach.

\subsection{Data and Measures}

We build our dataset from two publicly available tables of Wikimedia history published by the Wikimedia Foundation (WMF).\footnote{\url{https://wikitech.wikimedia.org/wiki/Analytics/Data\_Lake/Edits/Mediawiki\_history} (Archived: \url{https://perma.cc/CPM6-PY6F}; \url{https://dumps.wikimedia.org/other/mediawiki\_history/readme.html} (Archived: \url{https://perma.cc/3DDJ-9FXS})}
Although Wikipedia is published and collaborated on in several languages, the vast majority of knowledge regarding collaboration on Wikipedia is derived from studies of English Wikipedia \cite{hecht_tower_2010, hara_cross-cultural_2010}.  To support generalizability, we analyze data from  23 language editions of Wikipedia where edit quality flags are displayed in the RCFilters interface.
To ensure that we have variation in our outcomes, we exclude wikis with less than three edits above and below each threshold (see §\ref{sec:thresholds}) from each sub-analysis.
For all of our analyses, our unit of analysis is the \emph{revision}. Revisions correspond to a single edit to a page by a participant on Wikipedia.  We exclude revisions by bots since we care about how algorithmic flagging and social signals are used by human moderators.
Following guidance for RDDs \citep{lee_regression_2010}, we include only revisions very near to RCFilters thresholds, with ORES scores within 0.03 of the thresholds.

To manage the total size of our dataset, we analyze a sample that we construct by stratifying along several dimensions: Wikipedia language edition; user registration status (§\ref{sec:signal}); whether the editor has a user page or not (§\ref{sec:signal}); whether an edit was reverted in 2 hours, 48 hours, or 30 days; and whether the revert was controversial (§\ref{sec:controversial}).
Then, we sample 5000 edits from within unique combinations of the variables. If there are less than  5000 edits in the given strata, we include all of them.
We adjust for this stratification using sample weights throughout our analysis.
Since RCFilters was introduced to different wikis at different times, %
we sample edits during the period immediately following the introduction of ORES but weight our sample according to the number of edits to each wiki over the entire study period. 
The numbers of observations sampled at each threshold, from each Wiki, and for each model are available in the supplementary material.

\subsubsection{ORES scores and RCfilter thresholds}
\label{sec:thresholds}

The continuous forcing variable used in our RDD analysis is a score from the ORES algorithm described in §\ref{sec:empirical}. Scores range from 0 to 1 and reflect the predicted probability that a revision is damaging. Because the ORES system has been under continual development over time, we obtain ORES scores created at the times revisions were made from a log maintained by the WMF.
The treatments in our analysis are whether edits to Wikipedia are flagged by RCFilters. These flags are applied if, and only if, a score from ORES exceeds a threshold.
This use of thresholds at arbitrary operating points is a feature of most algorithmic flagging systems.
The intuition behind our RDD is that---after adjusting for small differences in quality associated with marginally higher or lower scores---edits with ORES scores immediately above and below an arbitrary threshold will be similarly likely to receive both first-order and controversial sanctions. Consequently, any discontinuous change in reverts at one of the thresholds used by RCFilters can be attributed to the flag.

RCFilters uses multiple thresholds corresponding to green, yellow, orange, and red flags. By default, only orange, and red flags are shown, but users can configure which colors to display. Green flags and filters are to help Wikipedia editors find good edits. 
Our analysis considers only red, orange, and yellow flags, which correspond to thresholds making different trade-offs between precision (the proportion of flagged edits that are truly damaging) and recall (the proportion of truly damaging edits that are flagged). The red flag is labeled ``very likely damaging'' and corresponds to a high precision threshold. Orange flags corresponds to a ``likely damaging'' label with greater recall but less precision. Edits with a yellow flag are ``maybe damaging'' with a high recall but lower precision.  
RCFilters' thresholds are truly arbitrary and have changed over time and across language editions in response to shifts in the precision and recall of ORES models and in response to community feedback.
We were able to collect data on threshold configuration, fully trained ORES models, code, and the precise time that changes were deployed in the WMF server admin log. We combined these data to identify the precise thresholds that were active for each revision in our dataset.

\subsubsection{Sanctions}

Our outcome variable for answering RQ1 must capture sanctioning in Wikipedia. Following a large body of other social computing research, we measure sanctions as identity reverts \citep[e.g.,][]{halfaker_dont_2011, halfaker_rise_2013, teblunthuis_revisiting_2018, piskorski_testing_2017}. Identity reverts occur when a user undoes another user's edit by restoring a page to an earlier state and are measured by comparing hashes of page revisions \citep{halfaker_dont_2011}. 

That said, identity reverts are an imperfect measure of sanctioning. It is also possible for an individual to ``self-revert'' by undoing their own edit. We therefore only treat a revision as reverted if it was undone, but not by a self-revert. We also limit our measure of sanctioning to revisions that are undone within 48 hour to avoid problems related to mass revert actions such as ``blanking'' of pages that result in false positives. We are confident that 48 hours is a reasonable window because most damage to Wikipedia will be undone within that amount of time \citep{geiger_when_2013} and a 48 hours window will include reverts caused by RCFilters since any effect of RCFilters is likely to occur quickly.

\subsubsection{Controversial sanctions}
\label{sec:controversial}

Our outcome variable for answering RQ2 and RQ3 measures controversial sanctions. We follow \citet{piskorski_testing_2017} by measuring controversial sanctions as identity reverts that are subsequently reverted by a third party.  Specifically, we label a sanction as controversial if the sanction is undone by a third editor who was not the original editor or the reverting editor.  Such interactions likely correspond to cases in which a third party observes the initial revert, disagrees with the initial sanction and then acts to reverse the sanction.

\subsubsection{Social signals}

\label{sec:signal}
Answering our RQ1 and RQ3 requires that we identify underprofiled and overprofiled individuals in our empirical setting. Drawing from research and documentation for Wikipedia moderators, we identify two such measures shown in the RCFilters interface shown in Figure \ref{fig:rcfilters}. 
Our first measure is whether an editor was logged into an account. Unregistered editors act on Wikipedia without logging in and registered contributors are those that edit with accounts. Because they are identified by their IP address rather than by a chosen username, unregistered editors are also referred to as ``IP editors'' or ``anons.'' Unregistered editors are associated with misbehavior and have long had a controversial status on Wikipedia \cite{mcdonald_privacy_2019}. Geiger and Ribes described how tools for moderators highlight unregistered editors \cite{geiger_work_2010}.
De Laat argued that unregistered editors on Wikipedia are overprofiled in that they are at higher risk to have their contributions rejected unfairly \citep{de_laat_use_2015, de_laat_profiling_2016}.

Second, the RCFilters interface indicates whether the editor has created a user page. User pages are Wikipedia's version of profile pages. Not having a user page is a social signal of newness because most committed users will create a user page early into their experience in Wikipedia \cite{ayers_how_2008}. The presence or absence of pages in Wikipedia is indicated with a subtle user interface clue: links to pages that do not exist are rendered in red, whereas links to pages that exist are blue. For example, Figure \ref{fig:rcfilters} shows the user ``Mashlova'' whose name is shown in red and would be identified as a newcomer.
De Laat cited the absence of a user page as a second example of an indicator of vandalism that will result in overprofiling \cite{de_laat_profiling_2016}. 
We measure whether an editor's user page exists at the time of a given contribution by matching the titles of user pages against the editor's username and checking if the creation of the user page was prior to the edit in question.  We only include registered editors in our analysis of overprofiling based on user pages.

\begin{table}
\centering
\footnotesize
\begin{subtable}{0.47\linewidth}
\begin{tabular}{llrr}
 Threshold & Edit type & N. & Prop. \\ 
  \hline
Maybe dam. & Not reverted & 12,403,717 & 0.87 \\ 
  Maybe dam. & Rev. controversial & 69,395 & 0.00 \\ 
  Maybe dam. & Rev. not cont. & 1,757,866 & 0.12 \\ 
  Maybe dam. & \textbf{Total} & 14,230,978 & 1.00 \\ 
   \hline
Likely dam. & Not reverted & 1,254,219 & 0.55 \\ 
  Likely dam. & Rev. controversial & 31,652 & 0.01 \\ 
  Likely dam. & Rev. not cont. & 1,009,108 & 0.44 \\ 
  Likely dam. & \textbf{Total} & 2,294,979 & 1.00 \\ 
   \hline
V. likely dam. & Not reverted & 58,474 & 0.15 \\ 
  V. likely dam. & Rev. controversial & 12,545 & 0.03 \\ 
  V. likely dam. & Rev. not cont. & 323,762 & 0.82 \\ 
  V. likely dam. & \textbf{Total} & 394,781 & 1.00 \\ 
   \hline
\end{tabular}

\caption{Counts and proportions of edits by whether an edit was reverted or controversially reverted in the neighborhood of each threshold.}\label{tab:edit.stats}
\end{subtable}
\qquad
\begin{subtable}{0.47\linewidth}
\begin{tabular}{llrr}
 Threshold & Editor type & N. & Prop. \\ 
  \hline
Maybe dam. & Reg. No User Page & 4,006,466 & 0.28 \\ 
  Maybe dam. & Reg. User Page & 3,797,451 & 0.27 \\ 
  Maybe dam. & Unregistered & 6,415,271 & 0.45 \\ 
  Maybe dam. & \textbf{Total} & 14,219,188 & 1.00 \\ 
   \hline
Likely dam. & Reg. No User Page & 281,964 & 0.12 \\ 
  Likely dam. & Reg. User Page & 26,459 & 0.01 \\ 
  Likely dam. & Unregistered & 1,982,985 & 0.87 \\ 
  Likely dam. & \textbf{Total} & 2,291,408 & 1.00 \\ 
   \hline
V. likely dam. & Reg. No User Page & 21,630 & 0.05 \\ 
  V. likely dam. & Reg. User Page & 687 & 0.00 \\ 
  V. likely dam. & Unregistered & 371,499 & 0.94 \\ 
  V. likely dam. & \textbf{Total} & 393,816 & 1.00 \\ 
   \hline
\end{tabular}

\caption{Counts and proportions of edits by whether an editor was registered or had a user page in the neighborhood of each threshold.}\label{tab:editor.stats}
\end{subtable}
\caption{Summary statistics from our full dataset. \label{tab:summary.stats}}
\end{table}

\section{Analytic plan}
\label{sec:analytic}

\newcounter{equationcnt}
\newcounter{figuretmp}
\setcounter{figuretmp}{\thefigure}
\setcounter{figure}{0}

Our analysis comprises Bayesian logistic regression models in two parallel analyses. 
The first analysis treats our dichotomous measure of whether edits are reverted as an outcome. This begins with an ``adoption check'' (§\ref{sec:adoption}) that describes the causal effects of flagging on reverts in general. The adoption check is a prerequisite to answering our research questions. The rest of the first analysis (§\ref{sec:results-rq1}) answers RQ1 by comparing the effect of RCFilters on edits by overprofiled users to its effect on other editors. 
Our second analysis is very similar but uses controversial reverts as the outcome, and analyzes only reverted edits to model the probability that a revert is controversial. It begins by  answering RQ2 (§\ref{sec:results-rq2}) in an analysis similar to the adoption check but with controversial sanctions as an outcome and with a dataset limited to overprofiled users. The rest of the second analysis (§\ref{sec:results-rq2}) answers RQ3 and is similar to RQ1 but with controversial reverts as the outcome in place of reverts.

Although our models use different sets of edits and outcomes, they all have the same logistic regression structure shown in Equation \ref{eq:model}. 

\begin{align}
    \mathrm{log}\left(\frac{P\left(Y_r\right)}{1-P\left(Y_r\right)}\right) &= 
     \alpha_{1}\left(score_{r} - c_{jw}\right) 
    + \tau_j \mathbf{1} \left[score_{r} > c_{jw}\right] \nonumber \\
    &+ \alpha_{2}\left(score_{r}-c_{jw}\right) \mathbf{1} \left[score_{r} > c_{jw}\right] + \alpha_w
\label{eq:model}
\end{align}

\noindent Our goal is to estimate $\tau_j$ which is the causal effect of being flagged at level $j$, where $j \in \{1,2,3\}$ corresponds to labels of ``maybe damaging,'' ``likely damaging'' and ``very likely damaging.'' For each cutoff on each wiki, we select revisions whose ORES scores are within a $\pm0.03$ window of the cutoff $(c_{jw})$.  
Following established approaches to RDD, we fit ``kink'' models that allow for a change in slope at the discontinuity \cite{lee_regression_2010, litschig_impact_2013}.  The slope before the discontinuity is $\alpha_1$ and the change in slope is $\alpha_2$. The indicator function is represented by $\mathbf{1}$.  Our models include fixed effects for wiki ($\alpha_w)$ to account for differences in the rates of sanctioning between wikis.

We use Bayesian inference to estimate our models for two reasons.  First, virtually all edits above the ``very damaging'' level are reverted in some of the wikis we analyze.  The presence of near-perfect ``separation'' creates estimation problems for classical numerical approaches \cite{allison_convergence_2004}. Preferred solutions to this problem in non-Bayesian frameworks include penalized likelihood methods that introduce bias.  Our Bayesian approach uses weakly-informative priors that are conservative but avoid the problem of separation. 
The second reason we use Bayesian inference is that it makes it easy to compare estimates across models. 
Prior work at CSCW by \citet{gan_gender_2018} used a similar rationale for adopting Bayesian logistic regression.
In Bayesian analysis, fitted models take the form of posterior distributions constituting a probability distribution of model coefficients conditional on our model, data and priors.  We consider a hypothesis supported if it is consistent with at least 95\% of posterior draws. In other words, we accept a given hypothesis if our parameter estimate has the predicted sign and the 95\% credible interval does not contain 0. This is the Bayesian analog to testing a hypothesis with $\alpha=0.05$.
We fit our models using the rstanarm package (version 2.19.3) and the default priors that are provided for reference in the supplementary material.

\section{Adoption Check}
\label{sec:adoption}

Before presenting results from hypothesis tests associated with our research questions, we first establish that RCFilters was adopted by Wikipedia moderators and that it had an effect on sanctioning behavior. This establishes a baseline necessary to answer RQ1 regarding the differential effects of RCFilters between overprofiled users and others.  Null effects in RQ1 might simply reflect that the system was not used. A successful adoption check rules out this possibility and sets up a credible null hypothesis test for RQ1.

We test the hypothesis that flagging increases the probability that an edit is reverted to demonstrate that RCFilters flags are being used by Wikipedia moderators.  Our estimates for $\tau_j$---as described in §\ref{sec:analytic}---should be positive if Wikipedia moderators are using flags in RCFilters to review potentially damaging edits.

\begin{figure}
  \centering
\begin{knitrout}
\definecolor{shadecolor}{rgb}{0.969, 0.969, 0.969}\color{fgcolor}
\includegraphics[width=\maxwidth]{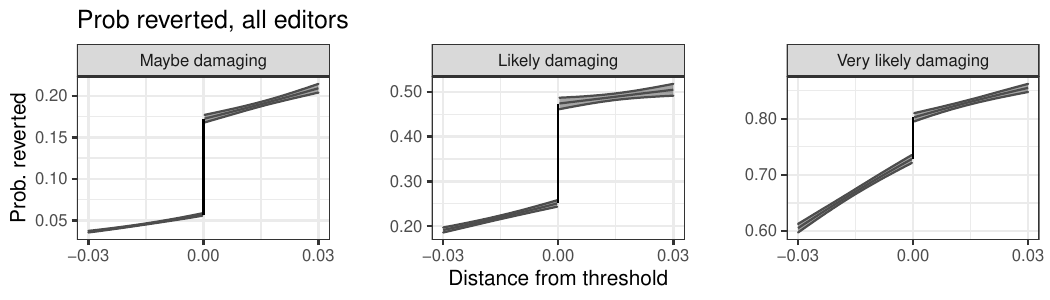} 

\end{knitrout}
\caption{Marginal effects plot showing model predicted relationship between ORES score and the probability that an edit will be reverted around the cutoffs for all contributors with 95\% credible intervals.\label{fig:adoption.me}}
\end{figure}

We find strong evidence that RCFilters was adopted and impacted sanctioning. Figure \ref{fig:adoption.me} visualizes this evidence: a marginal effects plot that illustrates our models' predicted likelihood of reverts across different ORES scores in the neighborhood of the thresholds. In each such plot, the $x$-axis shows the distance from the threshold such that discontinuities at 0 represent the effect of being flagged. The plots show modeled values for the English language edition of Wikipedia but are representative of relationships across all wikis.\footnote{Because intercepts are the only part of our model that depend on Wikis, slopes and the discontinuities caused by algorithmic flagging represent our inference over all our data.}
Figure \ref{fig:adoption.me} shows discontinuous increases in the likelihood of reversion at the ``maybe damaging'' and ``likely damaging'' thresholds in the left and center panels. 
We find the greatest effect at the ``maybe damaging'' threshold  ($\tau_1 = 1.23$ $[1.19;\allowbreak 1.28]$).\footnote{All $\tau$ parameter estimates are reported as log-odds ratios. The bracket notation indicates the 95\% credible interval.  In other words, the most likely value of the parameter is $1.23$, but there is a 95\% probability that the parameter lies in the interval $[1.19;\allowbreak 1.28]$.}
The effect at the ``very likely damaging'' threshold shown in the right-most panel is smaller  ($\tau_3 = 0.41$,  $[0.35;\allowbreak 0.46]$). 

The impacts of the ``maybe damaging'' and ``likely damaging'' flags on the likelihood of sanctioning are enormous. Figure \ref{fig:adoption.me} shows that likelihood of a revert for an edit just below the ``maybe damaging'' threshold is between 5.5\% and 5.8\%, indicating that reverts of unflagged edits are relatively rare. Being flagged 
with the ``maybe damaging'' flag causes a dramatic increase in the reversion probability to between 16.8\% and 17.7\% for edits just above the threshold.  
The effect of algorithmic flags at the ``likely damaging'' level is even more stark. We estimate that edits just below the ``likely damaging'' threshold are  likely to be reverted between 24.3\% and 25.8\% of the time, whereas similar edits just above the threshold are reverted between 46.1\% and 48.7\% of the time.  Being flagged at the ``very likely damaging'' threshold causes an increase in reversion probability from between 72.1\% and 73.5\% to 
between 79.5\% and 81\%.

\section{Results}
\subsection{RQ1: Effect of Flagging on Sanctioning}
\label{sec:results-rq1}

\begin{figure}
\centering

\begin{knitrout}
\definecolor{shadecolor}{rgb}{0.969, 0.969, 0.969}\color{fgcolor}
\includegraphics[width=0.7\linewidth]{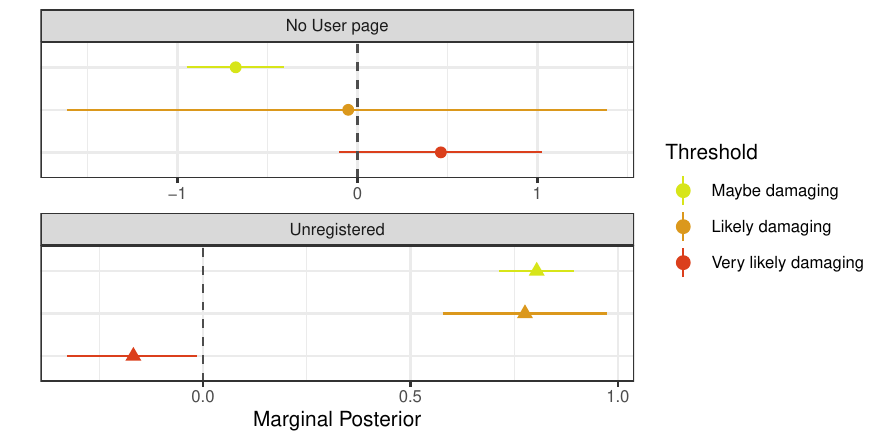} 

\end{knitrout}
\caption{Results for RQ1 showing point estimates and 95\% credible intervals for differences in the causal effect of flagging on sanctioning between overprofiled contributors and others.  A value greater than 0 indicates that our estimates of the effect for underprofiled contributors are greater than those for overprofiled contributors.
} %
\label{fig:h1.regplot}
\end{figure}

In our first research question (RQ1), we seek to understand how the increase in sanctioning caused by flagging 
affects discrimination against overprofiled users. If algorithmic flagging reduces overprofiling, as some computer scientists have argued \citep{kleinberg_human_2018}, the effect of flagging will be more scrutiny on users who are more likely to be given a pass. If algorithms simply reproduce discrimination, we will find no difference.
Results for hypothesis tests answering this question are shown in Figure \ref{fig:h1.regplot}, which visualizes the point estimates and credible intervals for differences in the causal effects of flagging on reverts between unregistered and registered contributors and between contributors with and without user pages. Values greater than 0 indicate that our estimated effect for the other users is greater than that for the overprofiled group. 

\begin{figure}
 \centering
\begin{knitrout}
\definecolor{shadecolor}{rgb}{0.969, 0.969, 0.969}\color{fgcolor}
\includegraphics[width=\textwidth]{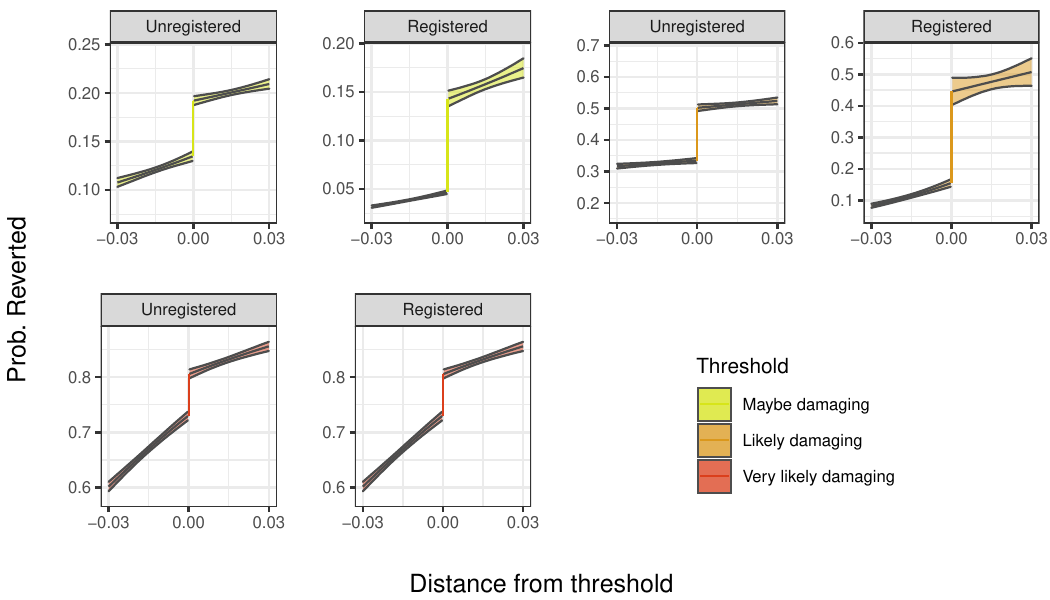} 

\end{knitrout}
  \caption{Results for RQ1 comparing unregistered and registered contributors are displayed in a marginal effects plot showing the model predicted relationship with 95\% credible intervals between ORES scores and reverts around the thresholds that trigger flags.  \label{fig:h1.me}}
\end{figure}

In support of the idea that algorithmic flagging can reduce overprofiling bias, we find that the overall effect of flagging is to increase demographic parity between registered and unregistered editors. Aggregating our posteriors over all three thresholds shows that the average effect over the three thresholds is greater for registered editors than for unregistered editors 
($\frac{1}{3}\sum_{j=1}^3{\tau^{\mathrm{Reg}}_j - \tau^{\mathrm{Unreg}}_j} =
0.45~[0.16;\allowbreak 0.6]$). 
The effect of flagging on reverts of registered editors is greater than the effect for unregistered editors at both the ``maybe damaging'' threshold ($\tau^{\mathrm{Unreg}}_1 - \tau^{\mathrm{Reg}}_1 = 0.8~[0.71;\allowbreak 0.89]$) and the ``likely damaging'' threshold ($\tau^{\mathrm{Unreg}}_2 - \tau^{\mathrm{Reg}}_2 = 0.78~[0.58;\allowbreak 0.97]$).
For an action by an unregistered contributor near to the ``maybe damaging'' threshold, being flagged increases the odds of being reverted by a factor of between 1.45 and 1.6 times. This is significantly less than the increase of between
3.16 and 3.68 times for registered contributors. 

However, at the ``very likely damaging'' threshold we find that the effects of flagging are stronger for unregistered editors than for registered editors ($\tau^{\mathrm{Reg}}_2 - \tau^{\mathrm{Unreg}}_2 = -0.17~[-0.33;\allowbreak -0.01]$). Being flagged increases the odds that an action  is reverted by a factor of between 1.43 and 1.62 times for an unregistered editor and by 1.11 and 1.49 times for registered contributors. However, as Table \ref{tab:summary.stats} shows, a far greater number of actions receive scores near to lower thresholds.  Thus, we focus on the lower thresholds in the following discussion.

Figure \ref{fig:h1.me} lets us interpret our models by making it possible to visually compare the effects of being flagged between overprofiled and underprofiled editors at a given threshold because the $y$-axes in each row span an identical range.  
The top-left panel shows how our models' linear predictions of how the probability of sanctioning for unregistered contributors  at the ``maybe damaging'' threshold jumps between 4.8 and 6.7
percentage points, from 13.5\% to 19.2\% on average. For registered editors, shown in the top-right of Figure \ref{fig:h1.me}, we estimate a jump of between 9.1 and 10.3 percentage points, from  4.6\% to 14.3\% on average. This is between
 3.3 and 4.6 percentage points greater than the jump for unregistered editors.
For unflagged edits that ORES scores near the ``maybe damaging'' threshold, an unflagged unregistered contributor has about the same odds of being sanctioned as a flagged registered contributor.

The bottom row of Figure \ref{fig:h1.me} shows that the change in sanctioning probability at the ``likely damaging'' threshold is
between 9.5 and 15.2 percentage
points greater for registered editors than for unregistered editors. 
For unregistered contributors, shown in the bottom-left of Figure \ref{fig:h1.me}, being flagged as ``likely damaging'' increases the probability of revert between  15  and 18.6 percentage points, from 33.5\% to 50.2\% on average. 
But for registered editors, shown in the bottom-right of Figure \ref{fig:h1.me}, we detect an even bigger jump of between 23.7  and 34.6 percentage points, from 15.5\% to 44.5\% on average.
For actions that ORES scores near the ``likely damaging'' threshold, unflagged actions by unregistered editors are far more likely to be reverted. Once flagged, actions by registered and unregistered editors are reverted at relatively similar rates.

These results show that flagging causes an increase in a decision system's demographic parity concerning registration status.  Actions by unregistered contributors that fall just above the cutoffs are much more likely to be reverted due to RCFilters---but the gap between actions by registered and unregistered contributors is much smaller when RCFilters has flagged an edit as ``maybe damaging'' or ``likely damaging.'' 
In this way, our analysis suggests that algorithmic flagging can reduce overprofiling bias.

\begin{figure}
 \centering
\begin{knitrout}
\definecolor{shadecolor}{rgb}{0.969, 0.969, 0.969}\color{fgcolor}
\includegraphics[width=\textwidth]{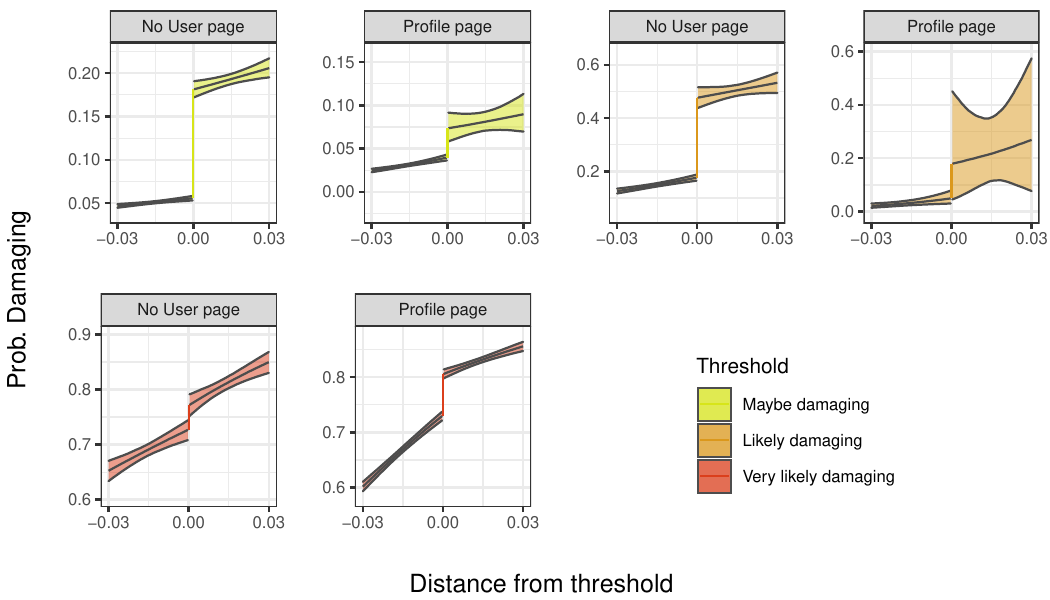} 

\end{knitrout}
  \caption{Results for RQ1 comparing contributors with and without user pages.
  Each panel shows a marginal effects plot with 95\% credible intervals of the modeled relationship between ORES scores and reverts around the thresholds that trigger flags.  \label{fig:h1.me.up}}
\end{figure}

Surprisingly, our results for our second measure of over-profiling in Wikipedia suggest a dynamic that is opposite in sign to the differences we observe between registered and unregistered editors at the 
``maybe damaging'' threshold ($\tau^{\mathrm{NoUP}}_1 - \tau^{\mathrm{UP}}_1 = -0.68~[-0.95;\allowbreak -0.41]$).  At the ``likely damaging''  ($\tau^{\mathrm{NoUP}}_2 - \tau^{\mathrm{UP}}_2 = -0.05~[-1.61;\allowbreak 1.39]$) and the ``very likely damaging'' ($\tau^{\mathrm{NoUP}}_2 - \tau^{\mathrm{UP}}_2 = 0.46~[-0.1;\allowbreak 1.03]$) thresholds, we do not detect  differences in effect size between contributors with and without user pages. 
At the ``maybe damaging'' threshold, we find that flagging increases the odds that an editor without a user page is reverted   between 3.47 and 4.06 times. This is significantly more than the increase of 
between 1.47 and 2.46 times for registered contributors. 

As above, we interpret these odds ratios using marginal effects plots shown in Figure \ref{fig:h1.me.up}.   The top-left plot in the figure shows our models' linear predictions of the probability of reverting for contributors without user pages near to the ``maybe damaging''  threshold.  For these editors, being flagged as ``maybe damaging'' increases the chances of sanctioning by 11.4 and 13.8
percentage points, from 5.6\% to 18.1\% on average. 
In the top-right of Figure \ref{fig:h1.me.up}, we see a jump of between 2.2 and 4.8 percentage points, from  4\% to 7.4\% on average for editors that have created user pages. This is between
 9.7 and 8.4 percentage points less than the jump for contributors without user pages.

\subsection{RQ2: Effect of flagging on controversial sanctioning}
\label{sec:results-rq2}

\begin{figure}
  \centering
\begin{subfigure}[t]{\textwidth}
  \centering
\begin{knitrout}
\definecolor{shadecolor}{rgb}{0.969, 0.969, 0.969}\color{fgcolor}
\includegraphics[width=0.7\linewidth]{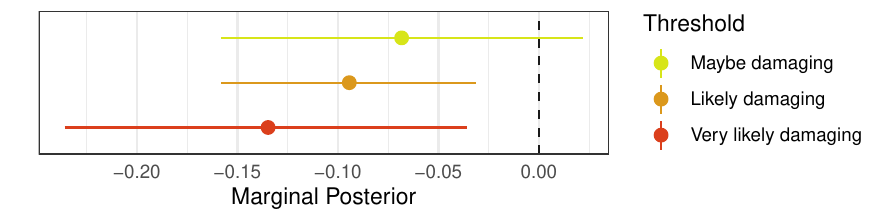} 

\end{knitrout}
\caption{Parameter estimates and 95\% credible intervals for the effects of flagging on  whether reverts are controversial for unregistered editors.  \label{fig:h2.regplot.anon}}
\end{subfigure}
~
\begin{subfigure}[b]{\textwidth}
\centering  

\includegraphics[width=1\linewidth]{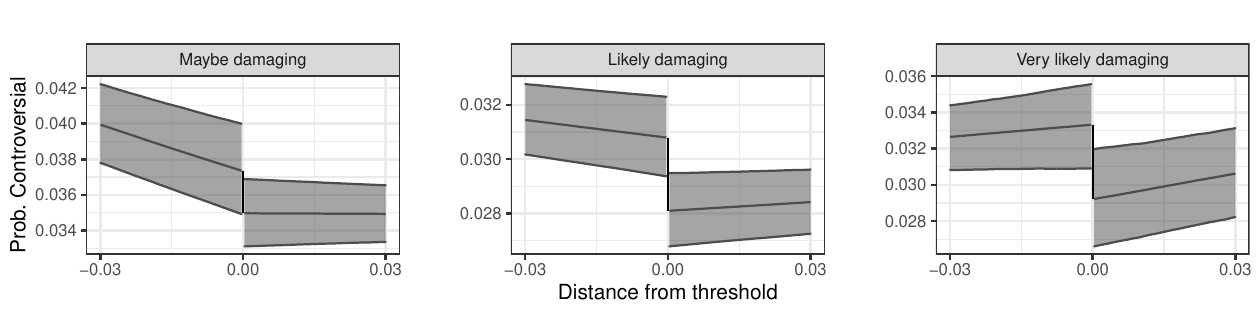} 

\caption[RQ2. me plot anon]{Marginal effects plots with 95\% credible intervals for models predicting whether a revert is controversial, for unregistered editors. \label{fig:h2.me.anon}}
\end{subfigure}
\caption[RQ2. plot anon]{Results for RQ2: flagging causes a small but detectable decrease in the likelihood that an action by an unregistered contributor receives a controversial sanction.}
\end{figure}

Consistent with the idea that algorithmic flagging can support fairness, we find that having an ORES score cross the ``likely damaging'' or ``very likely damaging'' thresholds decreases the chances that a revert will be controversial for unregistered editors.
These results are visualized in Figure \ref{fig:h2.regplot.anon}. We have less confidence in the effect at the ``maybe damaging'' threshold because our 95\% credible interval includes 0 ($\tau^{\mathrm{Unreg}}_{1}=-0.07;\allowbreak \mathrm{CI}=[-0.16;\allowbreak 0.02]$). 

\begin{figure}
  \centering
\begin{subfigure}[t]{\textwidth}
  \centering
\begin{knitrout}
\definecolor{shadecolor}{rgb}{0.969, 0.969, 0.969}\color{fgcolor}
\includegraphics[width=0.7\linewidth]{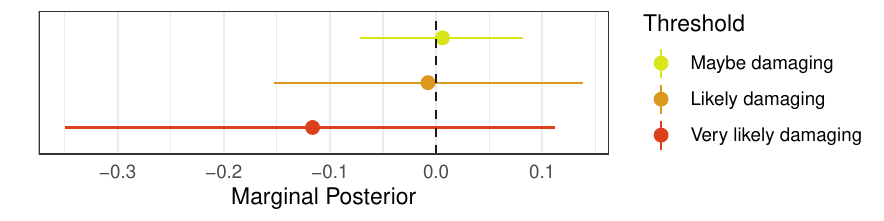} 

\end{knitrout}
\caption{Parameter estimates and 95\% credible intervals for effects of flagging on whether reverts are controversial for editors without user pages.}
\label{fig:h2.regplot.up}
\end{subfigure}
~
\begin{subfigure}[b]{\textwidth}
\centering  

\includegraphics[width=\textwidth]{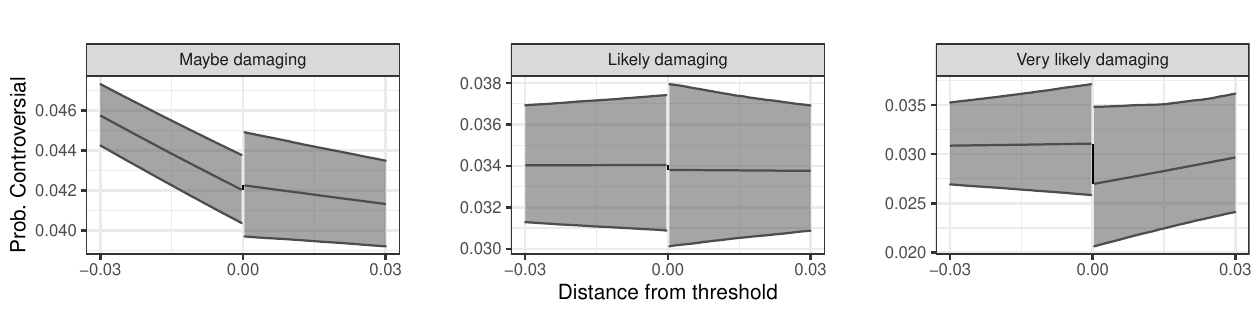} 

\caption[RQ2. me plot up]{Marginal effects plots with 95\% credible intervals for models predicting whether a revert is controversial, for contributors without user pages.}
\label{fig:h2.me.up}
\end{subfigure}
\caption{Results for RQ2 comparing contributors with user pages to those without show no detectable effect of flagging on controversial sanctioning.}
\end{figure}

We estimate that being flagged at the ``likely damaging'' level results in a change in the odds that a sanction is controversial by a factor between 0.85 and 0.97.  Figure \ref{fig:h2.me.anon} shows the modeled relationship between ORES scores and the probability of a controversial sanction in the neighborhood of the thresholds for English Wikipedia.  On the left plot, we see that being flagged changes unregistered contributor's likelihood of a controversial revert from a possible increase of 
0.27 percentage points to a possible decrease of
 0.55 percentage points, a change from 
3.08\% to 
2.81\%
on average.

We observe a similar effect of flagging at the ``very likely damaging'' threshold ($\tau^{\mathrm{Unreg}}_{2}=-0.13;\mathrm{CI}=[-0.24;\allowbreak -0.04]$): the odds that a revert is controversial are between 0.79 and 0.97 times smaller. On the right side of Figure \ref{fig:h2.me.anon}, we find that being flagged decreases the probability that a sanction to an action by an unregistered editor is controversial by between 0.11  and
0.89 percentage points, a change from 
from 3.33\% to 
2.92\%
on average.

However, we did not detect effects of flagging when the reverted editor lacks a user page at the ``maybe damaging'' ($\tau^{\mathrm{NoUP}}_{1}=0.01;\mathrm{CI}=[-0.07;\allowbreak 0.08]$), ``likely damaging'' ($\tau^{\mathrm{NoUP}}_{2}=-0.01;\mathrm{CI}=[-0.15;\allowbreak 0.14]$), or ``very likely damaging'' ($\tau^{\mathrm{NoUP}}_{3}=-0.12;\mathrm{CI}=[-0.35;\allowbreak 0.11]$) thresholds. 
Our results for RQ2 for unregistered editors show that flagging decreases the rate of controversial sanctions.  Although controversial sanctions do not precisely correspond to false-positive sanctions, we take this finding as evidence that flagging decreases the false positive rate of the decision system.
We address the inconsistencies between our results for unregistered editors and editors without user pages in our discussion (§\ref{sec:discussion}).

\subsection{RQ3: Social signals and effects of flagging on controversial sanctioning }
\label{sec:results-rq3}

\begin{figure}
\centering

\includegraphics[width=0.7\linewidth]{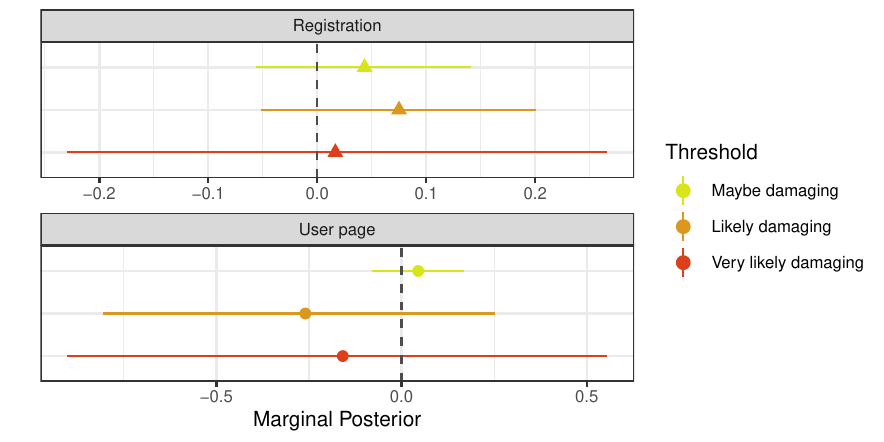} 

\caption{Results for RQ3 showing the difference in our parameter estimates between overprofiled editors and others with 95\% credible intervals. Values greater than 0 would indicate that the effect for underprofiled editors is greater than that for overprofiled editors.}
\label{fig:h3.reg.plot}
\end{figure}

To answer RQ3, we largely replicate the analysis conducted for RQ1 with the dependent variable used in RQ2. Results shown in Figure \ref{fig:h3.reg.plot} provide weak evidence that a decrease in controversial sanctioning may be greater for 
registered than for unregistered contributors at the ``maybe damaging'' ($\tau^{\mathrm{Reg}}_1 - \tau^{\mathrm{Unreg}}_1 = 0.04$ $[-0.06;\allowbreak 0.14]$), ``likely damaging'' ($\tau^{\mathrm{Reg}}_2 - \tau^{\mathrm{Unreg}}_2 = 0.07$ $[-0.05;\allowbreak 0.2]$),
and ``very likely damaging'' ($\tau^{\mathrm{Reg}}_3 - \tau^{\mathrm{Unreg}}_3 = 0.02$ $[-0.23;\allowbreak 0.27]$) thresholds. 
However, our evidence weakly suggests that the effect for contributors with user profiles is greater than those for without at the ``maybe damaging'' threshold
($\tau^{\mathrm{UP}}_1 - \tau^{\mathrm{NoUP}}_1 = 0.05$ $[-0.08;\allowbreak 0.17]$) but the opposite seems true at the ``likely damaging'' threshold ($\tau^{\mathrm{UP}}_2 - \tau^{\mathrm{NoUP}}_2 = -0.26$ $[-0.79;\allowbreak 0.26]$) and ``very likely damaging'' ($\tau^{\mathrm{UP}}_3 - \tau^{\mathrm{NoUP}}_3 = -0.16$ $[-0.9;\allowbreak 0.56]$) thresholds. None of these estimates are statistically significant at the 95\% level. 

\section{Threats to Validity}
\label{sec:threats}
Our results are subject to a range of threats to validity that pertain to our ability to make causal claims, rule out alternative explanations, and establish the generalizability of our findings. First, there are several threats to our ability to draw causal inferences that are common to RDDs.
Formally, RDDs model an outcome $Y$ as a function of a continuous ``forcing variable'' $Z$, other covariates, and a cutoff $c$ such that $Z>c$ determines treatment assignment.  In principle, treatment assignment conditional on $Z$ is ``as good as random'' under two assumptions: (1) that agents have at most limited control over $Z>c$, and (2) that the relationship between $Y$ and $Z$ is smooth \cite{lee_regression_2010}.
Although the assumptions required for causal inference are fundamentally unverifiable, we believe that our RDD provides relatively strong evidence of causal relationships between flagging and sanctioning.

Our treatment, being flagged in RCFilters, is an ideal candidate for an RDD from the perspective of assumption (1) because editors are unlikely to have much control over the scores that their edits receive.  Although attempts to evade sanction by specially crafting edits to evade algorithmic detection are hypothetically possible, the authors of ORES and RCFilters believe they are unrealistic and very unlikely to be widespread.
Assumption (2) would be violated if any unobserved treatments affect our outcomes at discrete levels of ORES scores. This is certainly possible because ORES makes scores available via a public API. Indeed, we are aware of bots that automatically revert edits triggered by the ``very damaging'' threshold on some of the Wikipedia language editions in our sample and therefore have more reason to doubt results at this threshold. Despite this threat, our conclusions regarding how algorithmic flagging shapes fairness are substantively similar whether we consider this threshold or not.
Although we identified one anti-vandalism tool---a system called Huggle discussed in §\ref{sec:discussion}---that collects ORES damaging scores, it uses ORES scores as one feature in its own algorithmic model and, by default, presents predictions from this model to users as a list of edits sorted in order of likelihood of vandalism. Given these facts, we believe that it is unlikely that Huggle users will drive discontinuities in the relationship between ORES scores and our outcomes. 

A limitation of RDD analysis is that it estimates effects for observations in the neighborhood of the cutoff and results may not generalize far away from the cutoff.  Compared with most RDD analysis, ours has the advantage of multiple different thresholds. Although our results for the ``likely damaging'' and ``maybe damaging'' thresholds are substantively similar, causal effects may diverge more at operating points we have not considered. Future work on algorithmic bias using RDD should consider that results may depend on the choice of operating points used as RDD cutoffs. 

An additional threat to validity is raised by the extent to which the ORES models encode biases concerning editors who are unregistered or without profile pages. 
To assess this threat, we analyzed the bias of ORES models for each wiki that had deployed the system on December 19\textsuperscript{th} 2020 using their human-labeled training data according to the \textit{conditional calibration} approach to evaluating model bias  \cite{mitchell_prediction-based_2020}.\footnote{We chose conditional calibration as our fairness metric because it does not depend on the choice of threshold. This simplifies the analysis of a decision system with multiple thresholds.} In our case, this involves comparing the rate of damaging edits predicted by the model to the true rates for each type of editor.
We find that ORES exhibits bias against both unregistered editors and editors without user pages but that the extent of bias against unregistered editors is much greater. These findings are opposite in sign to what we would expect if model bias were driving our results.
We present detailed results from this analysis in our online supplement.

Our study design is also limited in that we cannot present causal evidence of the impact of social signals. Although RCFilters's algorithmic flags are distributed in a quasi-experimental way, overprofiled status is not.
There are a range of possible systematic differences between overprofiled users and others that might be driving our results for RQ1 and RQ3.
For example, if damaging edits by contributors who are unregistered or lack user pages are more difficult for ORES to detect, that might drive our findings of a decrease in overprofiling for RQ1. %
Although we believe that this particular threat is unlikely because it would require that overprofiled contributors be systematically more sophisticated than others---something our experience with ORES suggests is unlikely---we cannot rule out either the specific threat or a range of other possibilities. 
A promising direction for future work might involve experiments or quasi-experiments that can jointly vary social signals and algorithmic flagging.

Additionally, system designers will likely want to know how overall rates of sanctioning and controversial sanctions change before and after a system such as RCFilters is launched. Unfortunately, our analysis cannot answer this question directly.
In preliminary work, we attempted to draw a statistical comparison between Wikipedia governance before and after the introduction of ORES but high temporal variation in sanctioning behaviors made this type of aggregate change difficult to measure. Future studies should organize with communities to conduct planned and principled field experiments to study the causal effects of introducing such systems in online communities using the model being pioneered by \citet{matias_civilservant_2018}. 

Finally, a set of largely unanswerable threats involves questions of generalizability across our measures and empirical contexts.  
Although our theory of interactions between algorithmic flags and social signals is general, and although we study RCFilters across 23 distinct communities, languages, and cultures, we study a single moderator tool on one platform.
We cannot claim that our findings generalize beyond the specific pool of communities that we study. 
Additionally, we have considered only a small subset of possible social signals that may be used in online community moderation.
Clearly, we also cannot claim that our settings are representative of moderation in online communities in general. 
Like most other empirical studies in social computing, we must sadly leave these questions for further research.

\section{Discussion}
\label{sec:discussion}

In the broadest strokes, our work is potentially good news for advocates of algorithmic flagging in social computing systems. It provides some evidence supporting the idea that algorithmic flagging can reduce discrimination in the form of overprofiling bias and that it can increase fairness. Our adoption check (§\ref{sec:adoption}) provides strong evidence that RCFilters drives behavior and our answers to RQ1 (§\ref{sec:results-rq1}) suggests that flagging can level the playing field by increasing decision system demographic parity between unregistered and registered Wikipedia editors.  Flagged edits by these contributors are reverted at similar rates, but unflagged edits of comparable quality by registered editors are reverted relatively infrequently.
More good news comes in the form of our answer to RQ2 (§\ref{sec:results-rq2}) that suggests that flagging is associated with a decrease in controversial sanctions among some overprofiled users and provides evidence that algorithmic flagging systems can help moderators more accurately issue sanctions.

When it comes to the details, however, the picture that emerges from our results is much more contingent and mixed. Our analysis used two different measures of overprofiling in Wikipedia but the pattern of our results diverged substantially between the two. The optimistic story about the effects of algorithmic flagging on overprofiled users only describes our results for unregistered Wikipedia users. Our evidence on overprofiled users without user pages is much weaker and points, in part, in the direction of algorithmic flagging increasing discrimination. Why do these results diverge? What do these divergent results mean for theory?

One possible explanation is that editors without user pages are, quite simply, not particularly overprofiled. Of the two social signals we consider, registration status attracts far more attention from academics and community members in discussions of Wikipedia vandalism \citep[e.g.,][]{hill_hidden_2020}.
Our analysis for RQ2, where we did not detect changes in controversial sanctions for editors without user pages, is also consistent with the notion that contributors without user pages may not be overprofiled. 
If algorithmic flagging systems help moderators more accurately issue sanctions by reducing overprofiling, then flagging would not decrease controversial sanctioning for editors that are not overprofiled. 
However, this alone does not explain why the effect for editors without profile pages was larger than for editors with them.

Our results might be explained if model bias against contributors without user pages means that the set of flagged edits from these users are less damaging than flagged edits by contributors who have profile pages.
As discussed in §\ref{sec:threats} and documented in our online supplement, ORES models are sometimes biased against contributors without user pages, but they are even more biased against anonymous contributors. 
Our results make sense if the overprofiling of anonymous editors outweighs model bias against them, but the reverse is true for editors without user pages.

It is also plausible that our mixed results are evidence that algorithmic flags will substitute for some social signals used in overprofiling while reinforcing others. Our study analyzes only two of many possible social signals that online community moderators might use. A better understanding of which signals drive sanctioning misbehavior can help explain if and when algorithmic triage systems can increase fairness.
Our results suggest that algorithmic flags can substitute for some social signals and reduce overprofiling in online community moderation. Our results also suggest that they might reinforce social signals, make overprofiling worse, or introduce new forms of unfairness through encoded bias. 
Unfortunately, outcomes resulting from myriad factors acting at once
are likely contingent on details of sociotechnical arrangements and difficult to know \textit{ex ante}.

Although RQ2 suggests that algorithmic flagging can increase fairness for overprofiled contributors, our null results for RQ3 mean that we could not detect a difference in this effect between overprofiled editors and others.  
Uncertainty in our models for RQ3 is high enough that parameter values consistent with a substantive average effect that is either positive or negative are plausible.
A null effect for RQ3 might also be explained if meta-norms and improved information are more important to controversial sanctioning than bias introduced by algorithmic flags or social signals acting as cues.

Our work has several important implications for designers of algorithmic flagging systems and sociotechnical systems.
Scholars of human computer interaction, science and technology studies, and the law have all called for analyses of algorithmic fairness to move beyond biases inherent in algorithms to consider the systemic and downstream effects of algorithms in use \cite{selbst_fairness_2019, stevenson_assessing_2017, zhu_value-sensitive_2018}. 
Ultimately, we recommend that operators of algorithmic flagging systems should continuously evaluate decision system fairness metrics and seek to improve them according to their values. In that the ORES model is, itself, biased against overprofiled users, our results suggest that evaluating the fairness of model predictions is only one piece of understanding how an algorithmic system shapes fairness in contexts such as online community moderation.

Future work should rigorously construct and critique decision system fairness criteria in terms of their consequences.
The algorithmic fairness literature often treats algorithmic predictions as equivalent to final decisions.
Our work shows that sociotechnical decision systems with humans in the loop face distinctive and contextually sensitive epistemic, ontological, and ethical questions about how decision system fairness should be defined or measured \cite{kleinberg_human_2018, selbst_fairness_2019}.

Decision system fairness is particularly important in open production communities such as Wikipedia because of the trade-offs between quality control and the essential tasks of supporting newcomers and encouraging contribution \cite{halfaker_rise_2013, morgan_tea_2013}.  
Past work has shown that increased quality control efforts correspond to a decrease in newcomer engagement and have hypothesized that one mechanism is increased scrutiny of newcomers \cite{halfaker_rise_2013, teblunthuis_revisiting_2018}.  Similarly, although blocking anonymous edits to wikis has shown been shown to cause a decrease in reverted edits, it also leads to a decrease in positive contributions \cite{hill_hidden_2020}.  While it may be intuitive to think about edits that get sanctioned as obvious vandalism, many of the edits flagged at the ``maybe damaging'' threshold are authored by well-meaning newcomers \cite{halfaker_rise_2013}.  There's a potentially high cost to sanctioning these low quality but well-intentioned contributions. We believe that our results point to the benefit of tracking changes in the rate of sanctions to sensitive groups of community members in order to assure that such well-meaning contributors are not being driven away.

There are also lessons to learned from the impressive degree with which RCFilters shapes behavior. 
Although the choice of operating points in algorithmic systems is often framed as purely about trading off precision and recall, our work demonstrates that these choices can have a range of other important consequences.  
Our disparate findings at the ``very likely damaging'' threshold for overprofiling based on registration status reveal that an algorithmic tool might improve fairness at a given operating point but decrease it at another.  
Although thresholds allowed us to explore the effects of flagging on sanctioning behavior, this arbitrary flagging of actions applied by RCFilters brought disproportionate attention to contributions just above the thresholds compared to contributions just below.  Designers should think about whether using thresholds to trigger flagging in moderation interfaces is a fair practice at all.  Our results show that this leads to sanctioning behavior that is, like the thresholds, arbitrary.

What types of designs might support quality control support models that scrutinize contributions in proportion to the likelihood that the contributions deserve to be sanctioned? We see some inspiration in Huggle, a counter-vandalism tool for Wikipedia which sorts actions by the likelihood that they are damaging.\footnote{See discussion in \cite{halfaker_snuggle:_2014}} Huggle users are encouraged to review the highest likelihood edits first and only move onto lower likelihood edits once those reviews are complete.  Such a user experience might increase efficiency and fairness by better concentrating moderator attention wherever it can have the greatest benefits.

\section{Conclusion}

As algorithmic flagging becomes more integrated into online community moderation, it is important to understand its effects and consequences on overprofiling and fairness. 
We use a regression discontinuity analysis of the RCFilters to find and sanction misbehavior by volunteers on Wikipedia to consider how the use of algorithmic flagging and social signals interact.
We find that by drawing moderator attention to misbehavior by registered participants, algorithmic flagging can reduce overprofiling in certain contexts.
We also find that algorithmic flagging can support fairness by decreasing controversial sanctions of unregistered contributors. Our results also suggest that the same system may have much less effect, and might even increase discrimination, for other types of overprofiled users.

Studies of machine learning in high-stakes settings like employment, education, and criminal justice trace how algorithms can encode discriminatory patterns in human behavior but might also improve fairness compared with human biases. Although the stakes are much lower, such questions are also pertinent to the use of machine predictions for online community moderation. We find that tools for predictive governance in a sociotechnical system can reduce overprofiling but  their effects are also difficult to anticipate.  %

Although our analysis of overprofiling based on registration status supports a rosy account of algorithmic flagging, our analysis of overprofiling based on user pages does not.  While contributors without user pages may be less overprofiled compared to unregistered contributors, our results also
suggest that the interaction between algorithmic flagging and social signals is complex and contingent. 
We suggest a need for future work that describes the kinds of social signals that are used in practice and explains how different types of information may be used alongside algorithmic flags. Finally, we present a methodological approach that we hope future studies of algorithmic tools in real-world sociotechnical systems might build upon to establish the causal effects of algorithmic systems without experimental intervention.

\begin{acks}

We are grateful to the anonymous CSCW reviewers and associate chairs for their keen insights and feedback.  We would also like to thank the Wikimedia Foundation for its support, members of the WMF analytics team including Andrew Otto, Luca Toscano, and Joal Allemandou for help with data access and computing infrastructure and members of the WMF research team including Jonathan Morgan for feedback early in project development. 
Thanks also go to members of the Community Data Science Collective who provided multiple rounds of feedback and contributed to copyediting including Kaylea Champion, Charles Kiene, Stefania Druga, Sohyeon Hwang, Jeremy Foote, and Aaron Shaw.
We also thank the WMF staff and volunteers who developed the systems we analyze including Roan Kattouw, the main developer of RCFilters, and the developers of ORES including Amir Sarabadani and Andy Craze. Special thanks to Amanda TeBlunthuis. Finally we owe an extra special thanks to the Wikipedia contributors whose digital traces we analyze. 
Portions of this work were facilitated through the use of advanced computational, storage, and networking infrastructure provided by the Hyak supercomputer system at the University of Washington. Financial support for this work came from the Wikimedia Foundation, from the National Science Foundation graduate research fellowship program \#2016220885, and from the University of Washington.

\end{acks}

\section*{Data Access}
A replication dataset including ORES scores, thresholds, and our sample of Wikipedia revisions, along with all of our code has been placed in the Harvard Dataverse archive and is available at the following URL: \url{https://doi.org/10.7910/DVN/E0RYJ4}

\received{June 2020}
\received[revised]{October 2020}
\received[accepted]{December 2020}

\bibliographystyle{ACM-Reference-Format}
\bibliography{ms.bib}

\end{document}